\documentclass[twocolumn, superscriptaddress, prb, showpacs]{revtex4-1}

\usepackage{ifthen}
\usepackage{graphicx}
\usepackage{amsmath}
\usepackage{color}
\usepackage[sort&compress]{natbib}

\begin{document}
\bibliographystyle{plainnat}

\title{Analyzing the frequency shift of physiadsorbed\\ CO$_2$ in metal
organic framework materials}

\author{Yanpeng Yao} \affiliation{Department of Physics and Astronomy,
Rutgers University, Piscataway, New Jersey 08854, USA}

\author{Nour Nijem} \affiliation{Department of Materials Science and
Engineering, University of Texas at Dallas, Richardson, Texas 75080,
USA}

\author{Jing Li} \affiliation{Department of Chemistry and Chemical
Biology, Rutgers University, Piscataway, New Jersey 08854, USA}

\author{Yves J. Chabal} \affiliation{Department of Materials Science and
Engineering, University of Texas at Dallas, Richardson, Texas 75080,
USA}

\author{David C. Langreth} \affiliation{Department of Physics and
Astronomy, Rutgers University, Piscataway, New Jersey 08854, USA}

\author{T. Thonhauser} \affiliation{Department of Physics, Wake Forest
University, Winston-Salem, North Carolina 27109, USA}

\date{\today}

\begin{abstract}
Combining first-principles density functional theory simulations with IR
and Raman experiments, we determine the frequency shift of vibrational
modes of CO$_2$ when physiadsorbed in the iso-structural metal organic
framework materials Mg-MOF74 and Zn-MOF74. Surprisingly, we find that
the resulting change in shift is rather different for these two systems
and we elucidate possible reasons. We explicitly consider three factors
responsible for the frequency shift through physiabsorption, namely (i)
the change in the molecule length, (ii) the asymmetric distortion of the
CO$_2$ molecule, and (iii) the direct influence of the metal center. The
influence of each factor is evaluated separately through different
geometry considerations, providing a fundamental understanding of the
frequency shifts observed experimentally.
\end{abstract}

%%%%%%%%%%%%%%%%%%%%%%%%%%%%%%%%%%%%%%%%%%%%%%%%%%%%%%%%%%%%%%%%%%%%%%%%
% 68.43.Bc   Ab initio calculations of absorbate structure and reactions
% 68.43.Fg   absorbate structure (binding sites, geometry)
% 84.60.Ve   Energy storage systems, including capacitor banks
%%%%%%%%%%%%%%%%%%%%%%%%%%%%%%%%%%%%%%%%%%%%%%%%%%%%%%%%%%%%%%%%%%%%%%%%
\maketitle

%%%%%%%%%%%%%%%%%%%%%%%%%%%%%%%%%%%%%%%%%%%%%%%%%%%%%%%%%%%%%%%%%%%%%%%%
\section{Introduction}
%%%%%%%%%%%%%%%%%%%%%%%%%%%%%%%%%%%%%%%%%%%%%%%%%%%%%%%%%%%%%%%%%%%%%%%%

physiabsorption of small molecules in multi-porous materials such as
zeolites and metal organic framework (MOF) materials has experienced a
surge of interest due to its potential for hydrogen-storage and
gas-separation applications.\citep{Zeolitic, Morris, Sholl, Rosi,
Rowsell, YunLiu, HughesMOF5, Caskey, HoffmannBreath, Valenzano,
HWuPCL10, bpdc, Nour_H2inter, Kong_rvH, Kong_ted, HWuMetal, WZhouMetal,
XiangMe, SunMetal, SumidaMgMOF74, DietzelNimof74, ChavanNiMOF74,
FitzZnmof74, Mgcell, FitzMOF5, synth1, WeiZ, synth3, synth4, synth5} MOF
structures have been widely investigated in order to find faster
absorption and higher storage densities, as well as proper binding
energies.\citep{Rosi, Rowsell, YunLiu, HughesMOF5, HoffmannBreath,
Caskey, Valenzano, HWuPCL10, bpdc, Nour_H2inter, Kong_rvH, Kong_ted,
HWuMetal, WZhouMetal, XiangMe, SunMetal, SumidaMgMOF74, DietzelNimof74,
ChavanNiMOF74, FitzZnmof74, Mgcell, FitzMOF5, synth1, WeiZ, synth3,
synth4, synth5} In order to design new MOFs with improved properties, it
is of critical importance to understand the nature of the interaction
between the absorbed molecule and the MOF host. Such understanding can
either be gained through theory, using first-principles
simulations,\citep{HoffmannBreath, Valenzano, HWuPCL10, bpdc,
Nour_H2inter, Kong_rvH, Kong_ted, HWuMetal, WZhouMetal, XiangMe,
SunMetal} or through experimental probes, such as infrared (IR)
absorption and Raman scattering.\citep{HoffmannBreath, Valenzano,
HWuPCL10, bpdc, Nour_H2inter, DietzelNimof74, ChavanNiMOF74,
FitzZnmof74,FitzMOF5, synth1}

Much progress has been made in improving the properties of MOFs. For
example, it has been shown that using unsaturated metal centers, such as
MOF74 with open Mg, Mn, Zn, Ni, Cu, or HKUST-1 with Cu, results in
higher absorption density for hydrogen and faster absorption at small
partial CO$_2$ pressures, the latter of which is highly desirable for
CO$_2$ capturing applications. \citep{Rowsell, YunLiu, HWuPCL10, Caskey,
Nour_H2inter, HWuMetal, WZhouMetal, XiangMe, SunMetal, SumidaMgMOF74,
DietzelNimof74, ChavanNiMOF74, FitzZnmof74} It has further been shown
that iso-structural MOFs with different open metal centers can have very
different absorption rates at low pressure.\citep{Caskey} In particular,
the electronically similar metal centers Mg and Zn result in a much
faster CO$_2$ uptake rate in Mg-MOF74 than in Zn-MOF74 at a pressure
smaller than 0.1~atm. \citep{Caskey} However, when previous research
yields contradictory results, it becomes difficult to gain further
insight; e.g.\ while the frequency shift of the asymmetric stretch mode
of absorbed CO$_2$ in Mg-MOF74 has been reported to be blue shifted in
one work using IR spectra and B3LYP-D$^*$ calculation,\citep{Valenzano}
it was reported as a red shift in another work using density functional
theory (DFT) with local density approximation (LDA)
simulations.\citep{HWuPCL10} Furthermore, a clear correlation between the
frequency shifts of the absorbed molecules and other absorption
properties such as the binding energy or the adsorption site is still
missing, which makes it difficult to directly correlate the observed
results with the physical nature of the absorption process.

Van der Waals density functional theory\citep{vdW-DF, potential_PRB,
vdW-DF2} (vdW-DF2) can be used as a very effective tool to understand the
molecule/MOF interactions. Unlike previous simulations
using LDA or GGA \citep{HWuPCL10} where long-range van der Waals
interactions are not included consistently, or B3LYP-D$^*$ \citep{Valenzano} where
the empirical parameters are used to incorporate the long-range
dispersion terms, in our vdW-DF2 method, the exchange-correlation
functional includes the---for these systems so important---long-range
van der Waals interactions between the MOF structure and the
physiadsorbed CO$_2$ molecules in a  seamless\citep{vdW-DF} and fully
self consistent way. \citep{potential_PRB} vdW-DF2 and its predecessor
vdW-DF have been successfully applied to many van der Waals
systems,\citep{Langreth_rev} ranging from simple dimers \citep{dimers} and
physiadsorbed molecules \citep{physisorption} to DNA and drug
design.\citep{drug} In particular, it has been demonstrated that vdW-DF2
can correctly capture the interaction and determine the adsorption
sites, binding energy, and vibrational frequencies for small molecules
absorbed in different MOFs.\citep{Kong_rvH,bpdc,Kong_ted,Nour_H2inter}

To resolve the contradicting results in literature, and to achieve a
better understanding of the physics that determines the frequency shift
of the absorbed molecule, in this work we combine both theoretical
first-principles electronic-structure simulation using vdW-DF2 and
experimental IR and Raman spectroscopy to study the CO$_2$ absorption in
the iso-structural MOFs Mg-MOF74 and Zn-MOF74. In the first step, we use
our experimental probes to determine the absorption behavior and the
vibrational frequencies of CO$_2$ absorbed in these MOFs. Then, in the
next step, we use vdW-DF2 to calculate the corresponding shifts and
determine the mechanisms that cause them.  Unlike in experiments,
first-principles simulations allow us to artificially freeze different
degrees of freedom of the system, enabling us to understand the
importance of different physical contributions to the CO$_2$ frequency
shifts. By analyzing the CO$_2$ asymmetric stretch frequency under
different geometries, we identify three different factors determining
the frequency shift of physiadsorbed CO$_2$, namely, the length of the
molecule, the asymmetric distortion, and the metal center.

%%%%%%%%%%%%%%%%%%%%%%%%%%%%%%%%%%%%%%%%%%%%%%%%%%%%%%%%%%%%%%%%%%%%%%%%
\section{Method}
%%%%%%%%%%%%%%%%%%%%%%%%%%%%%%%%%%%%%%%%%%%%%%%%%%%%%%%%%%%%%%%%%%%%%%%%

%%%%%%%%%%%%%%%%%%%%%%%%%%%%%%%%%%%%%%%%%%%%%%%%%%%%%%%%%%%%%%%%%%%%%%%%
\subsection{Metal organic framework synthesis}

Mg-MOF74,\citep{synth3} Zn-MOF74,\citep{synth4} Co-MOF74,\citep{synth5} and
Ni-MOF74\citep{Caskey} were synthesized according to procedures described
in the literature;\citep{Nour_H2inter, synth1}

\begin{description} \item[Mg-MOF74] 2,5-Dihydroxyterephthalic acid
(H$_2$DHB DC) (99 mg, 0.5 mmol) and Mg(NO$_3$)$_2$$\cdot$6H$_2$O
(257~mg, 1.0~mmol) were dissolved in the mixture of THF (7~mL), 1M NaOH
solution (2~mL), and water (3~mL) with stirring. The mixture was then
sealed in a Teflon-lined autoclave and heated at 110$^\circ$C for 3
days. The product was collected by filtration as a light-yellow
substance. Yield: 115~mg, 83\%.

\item[Zn-MOF74] The preparation of Zn-MOF74 was similar to that of
Mg-MOF74 except that Zn(NO$_3$)$_2$$\cdot$6H$_2$O (298~mg, 1.0~mmol) was
used instead of Mg(NO$_3$)$_2$$\cdot$ 6H$_2$O. Yield: 160~mg, 87\%.

\item[Co-MOF74] H$_2$DHBDC (150~mg, 0.75~mmol) and
[Co(NO$_3$)$_2$]$\cdot$6H$_2$O (186~mg, 0.75~mmol) were dissolved in
15~mL of THF-H$_2$O solution (50:50, v:v) with stirring. The mixture was
transferred to a Teflon-lined autoclave, which was then sealed and
heated at 110$^\circ$C for 3 days. Brown-red rod-shape crystals were
isolated by filtration and dried under vacuum. Yield: 130~mg, 50\%.

\item[Ni-MOF74] A mixture of H$_2$DHBDC (60~mg, 0.3~mmol),
[Ni(NO$_3$)$_2$]$\cdot$6H$_2$O (174~mg, 0.6~mmol), DMF (9~mL), and
H$_2$O (1~mL) was transferred to a Teflon-lined autoclave and heated in
an oven at 100$^\circ$C for 3 days. Brown crystalline powder was
collected by filtration and dried under vacuum. Yield: 75~mg, 72\%.
\end{description}

\noindent All as-synthesized materials were exchanged with fresh
methanol four times in a duration of 4 days, followed by drying in an
vacuum oven at room temperature, and annealing at 480~K overnight under
high vacuum before spectroscopic measurements.

%%%%%%%%%%%%%%%%%%%%%%%%%%%%%%%%%%%%%%%%%%%%%%%%%%%%%%%%%%%%%%%%%%%%%%%%
\subsection{IR and Raman spectroscopy}

IR absorption spectroscopy of CO$_2$ absorption in the MOFs was
performed in transmission at room temperature using a
liquid-N$_2$-cooled InSb detector. Approximately 12 mg of MOF powder was
pressed on a KBr support and mounted in a high-temperature high-pressure
cell (Specac product P/N 5850c) and heated to 200$^\circ$C in vacuum
(100 mTorr) overnight for complete desolvation. MOF74 samples were
activated by solvent exchange in methanol and drying in vacuum at room
temperature.  Subtraction of the gas phase CO$_2$ contribution to the IR
spectra was performed as described in Ref.~\onlinecite{bpdc}.

Raman spectroscopy measurements were performed using a solid state 532
nm laser. The activated sample was loaded into a Linkam FTIR
cooling/heating stage, and the sample was heated to 120$^\circ$C in
vacuum (900~mTorr) for complete dehydration. A laser power of
0.113--1.23~mWatt was used to avoid sample burning from the laser.

%%%%%%%%%%%%%%%%%%%%%%%%%%%%%%%%%%%%%%%%%%%%%%%%%%%%%%%%%%%%%%%%%%%%%%%%
\subsection{First-principles calculations}

For our first-principles calculations we used DFT as implemented in
$ABINIT$,\citep{ABINIT1,ABINIT2} utilizing vdW-DF2 to describe exchange and
correlation effects.\citep{vdW-DF2} Troulier-Martin type norm-conserving
pseudopotentials and a plane-wave basis are used,
\citep{Troulier} where the Zn 3$d$ semicore electrons are also considered
as valence electrons.  To ensure a full convergence of the structure and
energy, a kinetic energy cutoff of 45 Hartree is used for the plane-wave
basis. For structural relaxation, we start from the experimental atomic
positions and relax them using vdW-DF2 until the force on each atom is
less than $0.05$~eV/\AA.  The unit cell parameters are fixed to the
experimental values, where for the hexagonal unit cell
$a=25.881$~\AA and $c=6.8789$~\AA\ for Mg-MOF74\citep{Mgcell} and
$a=25.887$~\AA and $c=6.816$~\AA\ for Zn-MOF74,\citep{YunLiu} respectively.

Due to the complex interaction between CO$_2$ and MOF74, many local
minimum energy sites exist, making it difficult to identify the low-energy adsorption site, which corresponds to the low pressure absorption.
Indeed, multiple adsorption sites and binding energies are determined
from our simulations. In this work, we performed calculations with
multiple initial positions and orientations of CO$_2$ relative to the
MOF74, and choose the  site with the lowest total energy as the one for
our analysis. To guarantee that CO$_2$ is absorbed at the equivalent
lowest energy site in both Mg-MOF74 and Zn-MOF74, we start the atomic
position relaxation of CO$_2$ in Mg-MOF74 with the coordinates of the
relaxed CO$_2$ in Zn-MOF74, which is possible due to the isostructure
and symmetry of the two MOF systems. This ensures that we are comparing
the effect of the open metal site on CO$_2$ in the two different systems
on the same footing.

\begin{figure*}
\begin{center}
\includegraphics[width=11cm]{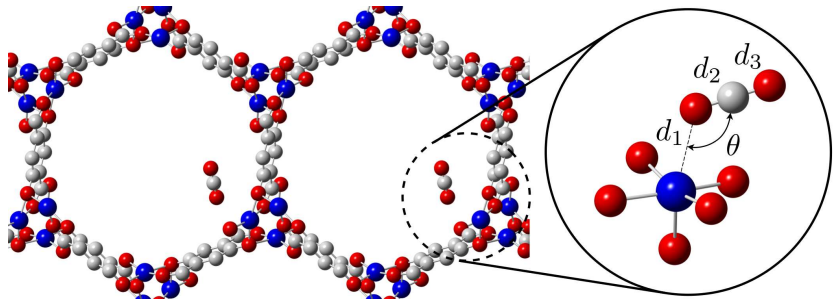}
\end{center}
\caption{\label{fig:MOF74}(Color online)Illustration figure of CO$_2$ absorbed in Zn-MOF74.
Calculated values of $d_1$, $d_2$, $d_3$, and $\theta$ for CO$_2$ in
Zn-MOF74 and Mg-MOF74 can be found in
Table~\ref{binding}.}
\end{figure*}

To calculate the frequency shifts we use a frozen-phonon approach in the
CO$_2$ molecules, where the MOF74 atoms are kept fixed, while each of
the atoms in the CO$_2$ is distorted in $\pm dx, \pm dy, \pm dz$
directions by small distortions $\Delta r=0.02$ \AA, to calculate the
force on each atom.  The symmetrized force matrix $2F(\Delta r)=F(\Delta
r)-F(-\Delta r)$ is constructed for only the atoms within the CO$_2$
molecule, based on the approximation that the interaction between the
CO$_2$ molecule and the MOF is weak and thus the effect of the vibration
of the MOF74 atoms on the CO$_2$ frequencies are negligible. However,
notice that, while the MOF74 atoms are kept fixed, the van der Waals
forces experienced by the CO$_2$ due to the presence of the MOF74 are
included by our vdW-DF2 calculation. To evaluate the effects of the surrounding MOF lattice, tests have been performed to allow
vibrations of several MOF74 atoms in the vicinity of the absorbed CO$_2$
molecule; frozen phonon calculations with such an extended force matrix
show no observable effect on the CO$_2$ frequencies. In general, our
studies show that calculated frequencies are converged to within less
than 1~cm$^{-1}$.

%%%%%%%%%%%%%%%%%%%%%%%%%%%%%%%%%%%%%%%%%%%%%%%%%%%%%%%%%%%%%%%%%%%%%%%%
\section{Results and discussion}
%%%%%%%%%%%%%%%%%%%%%%%%%%%%%%%%%%%%%%%%%%%%%%%%%%%%%%%%%%%%%%%%%%%%%%%%

%%%%%%%%%%%%%%%%%%%%%%%%%%%%%%%%%%%%%%%%%%%%%%%%%%%%%%%%%%%%%%%%%%%%%%%%
\subsection{Structure and binding energy}

\begin{table}
\caption{\label{binding}Calculated binding energy $\Delta E$ (kJ/mol),
angle $\theta$~(deg), and various distances $d_i$ and $l$ (\AA)
(where $l=d_2+d_3$) of the CO$_2$ molecule physiadsorbed in the MOF ;
see Fig.~\ref{fig:MOF74} for further details. The calculated free CO$_2$
parameters are $d_0=1.1630$ \AA\ and $l_0=2.3260$ \AA.}
\begin{tabular*}{\columnwidth}{@{\extracolsep{\fill}}lcccccr@{}}\hline\hline
MOF      & $\Delta E$ & $\theta$ & $d_1$ & $d_2-d_0$ & $d_3-d_0$ & $l-l_0$\\\hline
Mg-MOF74 & 35.4       & 120.8      & 2.53  & +0.0062   & --0.0059  & +0.0003\\
Zn-MOF74 & 26.9       & 116.4      & 2.69  & +0.0048   & --0.0039  & +0.0009\\\hline\hline
\end{tabular*}
\end{table}

The structure of CO$_2$ absorbed in MOF74 is illustrated in
Fig.~\ref{fig:MOF74}. In this work, we consider only CO$_2$ absorption
under low pressure smaller than 7 Torr, where the open metal site in
MOF74 is far from fully occupied. In our first-principles simulations
one CO$_2$ per six metal sites absorption is considered, corresponding
to the low-loading situation observed experimentally. For the binding
energy of CO$_2$ absorbed in Mg-MOF74, we find 35.4~kJ/mol, while it
binds somewhat weaker in Zn-MOF74 with 26.9~kJ/mol. Comparing the
distance of the CO$_2$ molecules with the open metal site, we find a
metal-oxygen distance ($d_1$ in Fig.~\ref{fig:MOF74}) of 2.53~\AA\ for Mg
versus 2.69~\AA\ for Zn. In other words, CO$_2$ binds stronger and closer
to the metal center in Mg-MOF74 than it does in Zn-MOF74.  At the same
time, similar metal-CO$_2$ angles ensure that the CO$_2$ is absorbed at
equivalent sites in both MOFs, with slightly different metal-CO$_2$
angles of 120.8$^\circ$ and 116.4$^\circ$ in Mg-MOF74 and Zn-MOF74,
respectively.

Further analyzing the structure of the absorbed CO$_2$ molecule, we see
that the C=O bond closer to the metal site ($d_2$ in
Fig.~\ref{fig:MOF74}) is elongated in both systems with a value of
1.1692~\AA\ in Mg-MOF74 and 1.1678~\AA\ in Zn-MOF74; in free CO$_2$ it
is 1.1630~\AA\ (denoted as $d_0$ in Table~\ref{binding}). On the other
hand, the C=O bond farther away from the metal center ($d_3$ in
Fig.~~\ref{fig:MOF74}) are both shortened, with a value 1.1571~\AA\ in
Mg-MOF74 and 1.1591~\AA\ in Zn-MOF74. A summary of distortions is shown
in Table~\ref{binding}. This distortion can be understood intuitively
via the interaction between the CO$_2$ and the metal center, where the
attraction from the metal center weakens (and thus elongates) the nearby
C=O bond, shortening the remaining C=O bond.  By comparing with the free
CO$_2$ value of 1.1630~\AA, we see that the CO$_2$ asymmetric distortion
in Mg-MOF74 is stronger than in Zn-MOF74.  Summing up the two C=O bonds
yields the overall length of the CO$_2$ molecule, which shows an
elongation of +0.0003~\AA\ and +0.0009~\AA\ for Mg-MOF74 and Zn-MOF74
absorption, as shown in Table~\ref{binding}.  It has been reported previously that CO$_2$ molecules absorbed in the MOF structure might experience some nonlinear distortion, i.e., an O-C-O angle differing from 180$^o$\citep{HWuPCL10}.  In this work, we also alow nonlinear distortion of the CO$_2$ molecule, and we find that the relaxed CO$_2$ molecule are only slightly bent after adsorbed within MOF74, featuring an O-C-O angle of $179.25^o$ in Zn-MOF74 and $178.97^o$ in Mg-MOF74.

As mentioned above, multiple local minimum adsorption sites are obtained
during the relaxation of CO$_2$ within the MOF74, leading to different
metal-CO$_2$ distances and binding energies. For instance, in Zn-MOF74
positions with similar location but Zn-CO$_2$ distances of 2.96~\AA\ and
3.09~\AA\ are found with binding energies of 25.6~kJ/mol and
24.8~kJ/mol, respectively.
%These sites might be possible adsorption
%sites for CO$_2$ at finite temperatures, where the thermal energy allows
%the CO$_2$ to move freely within these energetically and geometrically
%close sites.
In addition to these sites similar to the global minimum
metal site, we also find a secondary non-metal site for CO$_2$
absorption, which lies between two neighboring equivalent metal sites
along the direction of the MOF74 pore, with a distance of 3.99~\AA\ to
the nearest metal Zn site and a binding energy of 21.8~kJ/mol---clearly
much smaller than at the metal site. The distance between two equivalent
metal sites along the pore direction in MOF74 is the $c$ length of the
hexagonal unit cell, which is approximately  6.8~\AA, as mentioned
above. It is thus possible that at high CO$_2$ pressure, when all the
metal sites are occupied (1 CO$_2$ per 1 metal), more CO$_2$ can still
be absorbed within the MOF74 occupying the secondary site, with a
CO$_2$--CO$_2$ distance of approximately 3.4~\AA\ along the pore
direction, thus resulting in a higher (2 CO$_2$ per 1 metal) storage
density of CO$_2$ within MOF74.

%%%%%%%%%%%%%%%%%%%%%%%%%%%%%%%%%%%%%%%%%%%%%%%%%%%%%%%%%%%%%%%%%%%%%%%%
\subsection{CO$_2$ frequencies from experiment}

\begin{figure}[t]
\centering
\includegraphics[width=0.8\columnwidth]{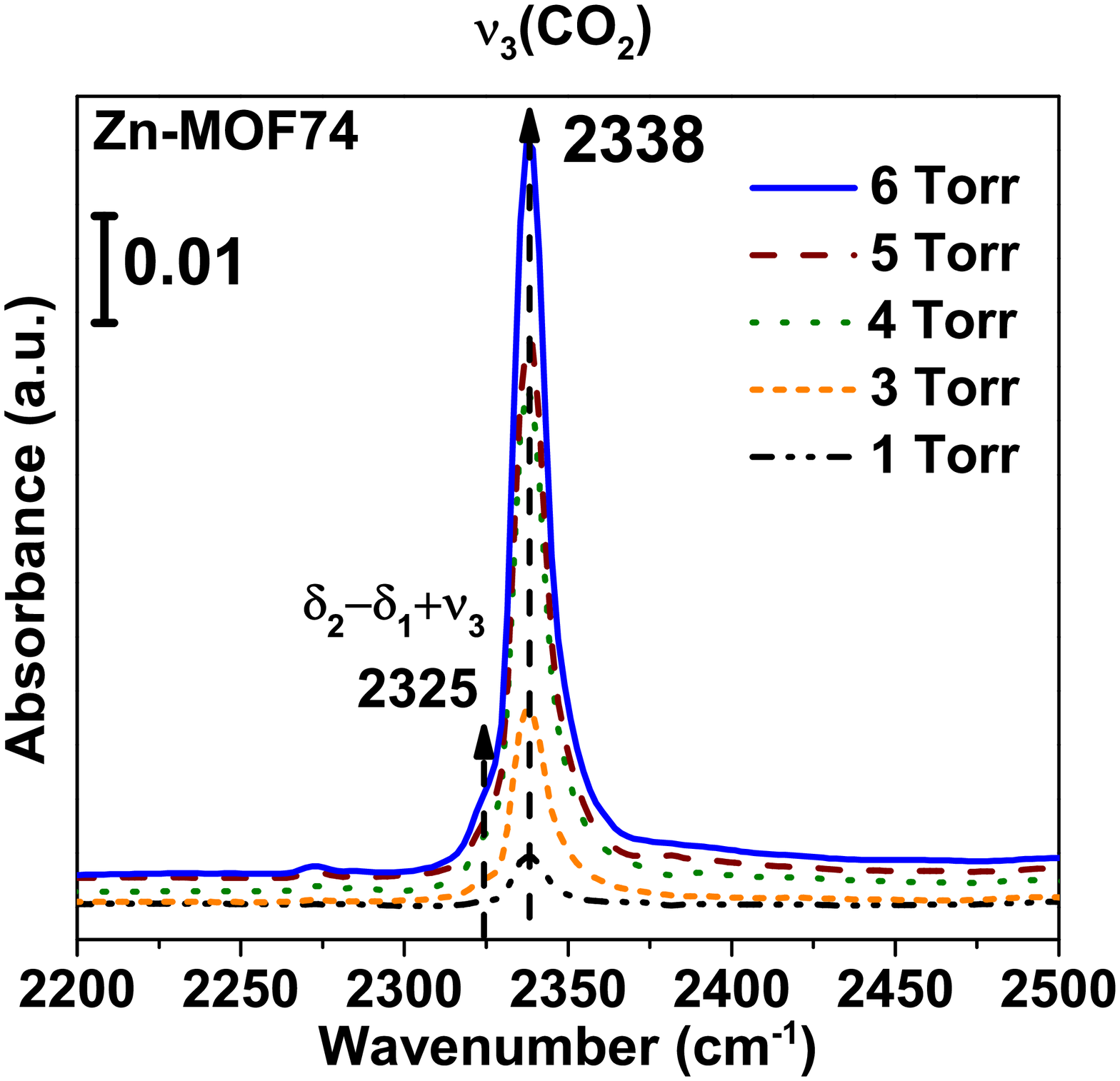}\\[4ex]
\includegraphics[width=0.8\columnwidth]{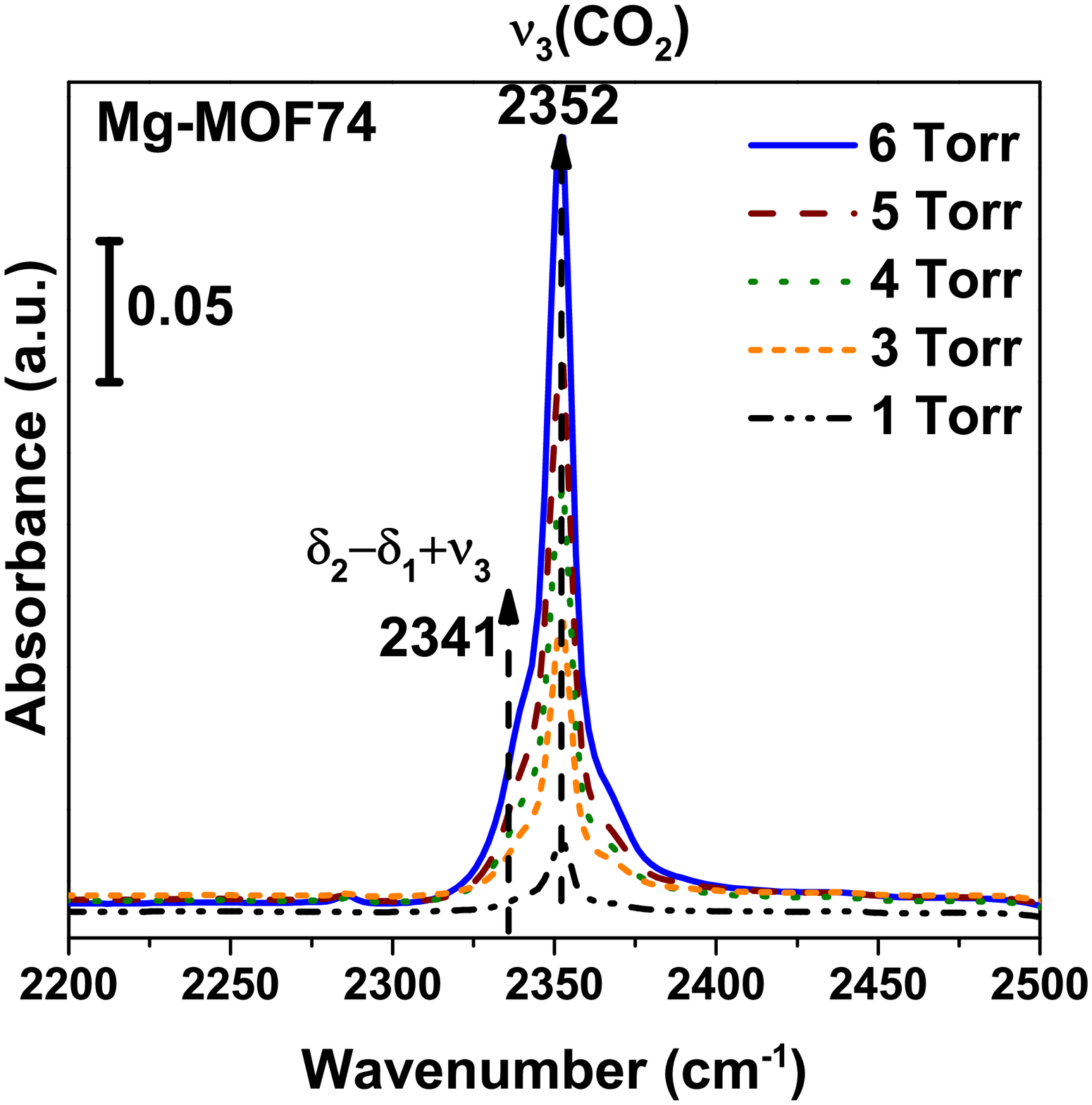}
\caption{\label{Asy}IR absorption spectra of CO$_2$ absorbed into
Zn-MOF74 (top) and in Mg-MOF74 (bottom) at changing CO$_2$ pressure
(1--6 Torr).}
\end{figure}

\begin{figure}
\centering
\includegraphics[width=0.8\columnwidth]{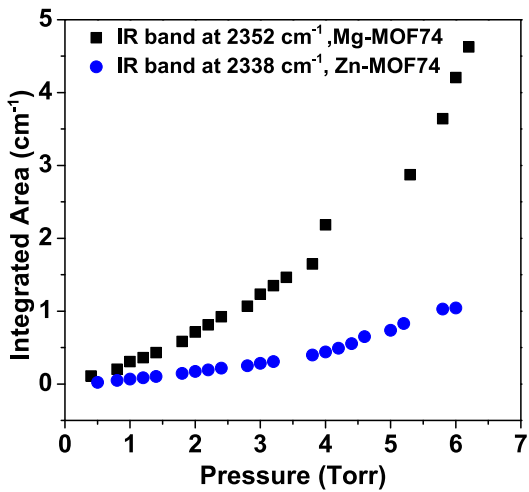}
\caption{\label{Inte}Integrated areas of IR absorption peaks for the
asymmetric stretch modes for CO$_2$ absorbed in Mg-MOF74 and Zn-MOF74 as
a function of CO$_2$ pressure.}
\end{figure}

IR absorption spectra of CO$_2$ absorbed in Mg-MOF74 and Zn-MOF74 are
shown in Fig.~\ref{Asy}, depicting a main IR absorption band attributed
to the asymmetric stretch of CO$_2$ at 2338 and 2352~cm$^{-1}$
for Zn-MOF74 and Mg-MOF74. The integrated areas of the IR band of the
asymmetric CO$_2$ stretch mode can serve as an indication of the
relative amount of CO$_2$ absorbed in MOF74. Figure \ref{Inte}
summarizes the integrated area as a function of CO$_2$ pressure for both
Mg-MOF74 and Zn-MOF74, showing that under the same pressure, the amount
of absorbed CO$_2$ is more for the case of Mg-MOF74 than it is for
Zn-MOF74. These results are consistent with isotherm measurements of
CO$_2$ reported in Ref.~\onlinecite{Caskey}.  Similar IR measurements
are also performed on Co-MOF74 and Ni-MOF74, showing shifts of the
asymmetric CO$_2$ stretch ($\sim$2340~cm$^{-1}$) similar to that in
Zn-MOF74.

In Fig.~\ref{Asy}, the shoulder peaks at 2325 and
2341~cm$^{-1}$ are attributed to the combination mode of the stretch
mode and the two non-degenerate bending modes (denoted as $\delta_1$ and
$\delta_2$ in Fig.~\ref{Asy}).  Experimentally, it is difficult to
identify the bending modes in the low frequency range because of their
weak intensity as compared to the MOF vibrations.  Therefore, we could
only resolve one of these non-degenerate bending modes at
$\sim$658~cm$^{-1}$.  The similarity of the bending mode for the Mg and
Zn cases is consistent with the close calculated values for both systems
shown in Table~\ref{frequency}.

Raman spectroscopy measurements were preformed to find the experimental
value of the symmetric stretch mode of absorbed CO$_2$ for the Co-MOF74
and Ni-MOF74. The symmetric stretch was found to be approximately
--1380 and --1382~cm$^{-1}$ for Co-MOF74 and Ni-MOF74,
respectively.  Because of the strong MOF Raman modes, the symmetric
stretch for Mg-MOF74 and Zn-MOF74 is not available. A summary of the IR
and Raman data for the frequencies of the absorbed CO$_2$ within MOF74
can be found in Table ~\ref{frequency}.

%%%%%%%%%%%%%%%%%%%%%%%%%%%%%%%%%%%%%%%%%%%%%%%%%%%%%%%%%%%%%%%%%%%%%%%%
\subsection{Analyzing the frequency shift}

When we consider the frequency shifts from the free CO$_2$ situation,
comparing with 2349~cm$^{-1}$ for free CO$_2$, the asymmetric stretch
for CO$_2$ absorbed in MOF74 has been shifted by $+3$ and
--11~cm$^{-1}$ for Mg and Zn-MOF74, respectively. Our first-principles
frozen-phonon calculations are in good agreement with this result and we
find the CO$_2$ asymmetric stretch mode experiences a slight redshift of
approximately --0.5~cm$^{-1}$ in Mg-MOF74 and --8.1~cm$^{-1}$ in
Zn-MOF74. Furthermore, the bending modes calculated also experience
redshifts of about --10~cm$^{-1}$ in both Mg-MOF74 and Zn-MOF74, which
is in excellent agreement with the experimentally observed --9~cm$^{-1}$
redshifts in these systems.  The calculated results for all the CO$_2$
modes are summarized in Table~\ref{frequency}.

We have shown above that CO$_2$ is binding stronger in Mg-MOF74, with a
closer metal-CO$_2$ distance and larger molecule distortions.  It is
thus puzzling to observe that the $\nu_3$ frequency shift for CO$_2$ in
Mg-MOF74 is much smaller compared with that in Zn-MOF74. There have been
many experiments and theoretical simulations studying the frequency
shifts of small molecules absorbed in MOFs, such as H$_2$, CO$_2$, CO,
etc.\citep{Valenzano, HWuPCL10, bpdc, Nour_H2inter, Kong_rvH, Kong_ted,
SumidaMgMOF74,  DietzelNimof74, synth1, ChavanNiMOF74, FitzZnmof74,
FitzMOF5} It is well known that the frequency shift of small molecule
vibrations has no obvious correlations with either the binding energy or
the adsorption site of the molecule within the MOF.  In our example here
with iso-structural Mg-MOF74 and Zn-MOF74, this behavior is even more
obvious, as Mg and Zn have similar electronic structures: the metal
centers have valence electrons of 3$s$ and 4$s$, respectively, except that
Zn has the additional fully occupied 3$d$ orbital electrons. Thus the
question arises as to what the physical reason is that determines the
frequency shift of the small molecules absorbed in the MOF.

\begin{table}
\caption{\label{frequency}Vibrational frequencies (cm$^{-1}$) of CO$_2$
physiadsorbed in MOF74.  Higher accuracy is kept for the calculated $\nu_3$ for further detailed analysis.}
\begin{tabular*}{\columnwidth}{@{\extracolsep{\fill}}llccr@{}}\hline\hline
       & system              & sym. $\nu_1$ & bend  $\nu_2$ & asym. $\nu_3$ \\\hline
       & free CO$_2^\dagger$ & 1388         & 667           &     2349      \\
       & Mg-MOF74            & ---          & 658           &     2352      \\
exp.   & Zn-MOF74            & ---          & 658           &     2338      \\
       & Co-MOF74            & 1380         & ---           &     2340      \\
       & Ni-MOF74            & 1382         & ---           &     2340      \\\hline
       & free CO$_2$         & 1298         & 646, 639       &     2288.5      \\
calc.  & Mg-MOF74            & 1294         & 636, 630       &     2288.0      \\
       & Zn-MOF74            & 1296         & 637, 632       &     2280.4      \\\hline\hline
\end{tabular*}
\raggedright$^\dagger$Taken from Ref.~\onlinecite{NIST}.
\end{table}

Several reasons may contribute to the frequency shift. For instance, by
analyzing the geometries of the absorbed CO$_2$, we noticed that the
absorbed CO$_2$ molecules have been distorted from their free molecule
geometry, exhibiting an off-center shift of the carbon atom, as well as
an elongated overall length of the molecule, as summarized in
Table~\ref{binding}(The slight nonlinear distortion of the CO$_2$ molecule is ignored here in this analysis). The change in the length of the molecule and the
asymmetric distortion can both affect the vibrational frequencies.
Beyond the effect of the change in the molecule geometry, the nearby
open metal center might also play a direct role in the frequency shift,
by attracting or repelling the nearby oxygen atom in the CO$_2$
molecules during the vibration.  To understand these different
contributions, in the following we analyze the influences of each of
these factors on the frequency of the CO$_2$ vibrations separately
and summarize their contributions in Table~\ref{shift}.

%%%%%%%%%%%%%%%%%%%%%%%%%%%%%%%%%%%%%%%%%%%%%%%%%%%%%%%%%%%%%%%%%%%%%%%%
\subsubsection{Change in molecule length}

We first analyze the effect of the changing molecule length. To do this,
we perform frozen-phonon calculations on the free CO$_2$ molecules fixed
to the lengths of the CO$_2$ absorbed in Mg-MOF (shown in the last
column of Table~\ref{binding}), while keeping the carbon atom at the
center of the molecule.  With this geometry, the CO$_2$ asymmetric
stretch is shifted by --1.6~cm$^{-1}$.  A similar calculation by setting
the CO$_2$ molecule length to the value of that absorbed in Zn-MOF74
yields a redshift in $\nu_3$ of --3.7~cm$^{-1}$.  Note that overall the
CO$_2$ in Mg-MOF74 is elongated by 0.0003~\AA, while that in Zn-MOF74 is
elongated by 0.0009~\AA.  The CO$_2$ with the longer length (in
Zn-MOF74) has a larger redshift of approximately --2.1~cm$^{-1}$ than
that in Mg-MOF74. That is, a longer molecule has more redshift, as
suggested by common sense.

%%%%%%%%%%%%%%%%%%%%%%%%%%%%%%%%%%%%%%%%%%%%%%%%%%%%%%%%%%%%%%%%%%%%%%%%
\subsubsection{Asymmetric distortion of the molecule}

Next, we take the asymmetric effect into consideration by placing the
CO$_2$ at exactly the same geometries as they are in the two MOFs, while
removing the surrounding MOFs. The frequency shift thus comes from the
change in both the length of the CO$_2$ molecule and the asymmetric
shift of the carbon atom. Our frozen-phonon calculations give an overall
frequency shift of \mbox{--0.5} and --3.0~cm$^{-1}$ for the
$\nu_3$ mode of CO$_2$ from Mg-MOF74 and Zn-MOF74, respectively.
Subtracting the previous results of only considering the length of the
molecule, the asymmetric distortion of the carbon atom causes a slight
blue shift in the asymmetric stretch with a value of 1.1 and
0.7~cm$^{-1}$ in Mg-MOF74 and Zn-MOF74.

To confirm these results, we also perform calculations using another
setup, where we keep the optimized CO$_2$ within the MOF and shift the
carbon atom to the center of the CO$_2$ molecule.  In this way, we
include the effect of the MOF environment and the length effect. The
only difference between this new setup and the optimized CO$_2$ in MOFs
is that the carbon atom off-center asymmetric effects are eliminated.
Our simulations show that the CO$_2$ in such a geometry has a frequency
shift of \mbox{--1.7} and --8.7~cm$^{-1}$ in Mg-MOF74 and
Zn-MOF74.  Comparing these results with the frequency shifts for relaxed
CO$_2$ in the same MOFs, we find similar differences of approximately
1.2 and 0.6~cm$^{-1}$ for the case of Mg-MOF74 and Zn-MOF74,
which are consistent with our previous results. It is thus clear that
the carbon off-center distortion indeed causes slight blue shifts in the
$\nu_3$ frequency. This result can also be understood in the following
way: by shifting the carbon atom off-center, the CO$_2$ atom is
distorted asymmetrically in a similar pattern as the asymmetric stretch
mode, which thus favors the asymmetric stretch. As shown in
Table~\ref{binding}, the asymmetric distortion in Mg-MOF74 is stronger
than that in Zn-MOF74, which is consistent with a larger blueshift
effect in the CO$_2$ $\nu_3$ frequency.

%%%%%%%%%%%%%%%%%%%%%%%%%%%%%%%%%%%%%%%%%%%%%%%%%%%%%%%%%%%%%%%%%%%%%%%%
\subsubsection{Effect of the metal center}

By comparing with the results for optimized CO$_2$ in the MOFs, the
previous results of the effects of the CO$_2$ molecule geometry change
on the $\nu_3$ frequency shifts yield the influence of the metal center
on the $\nu_3$ frequencies.  A direct comparison gives a frequency shift
of --5.1~cm$^{-1}$ for the Zn-MOF74 and 0~cm$^{-1}$ for Mg-MOF74.  This
result is quite a surprise, claiming that the open Zn center has a
strong redshift influence on the $\nu_3$ frequency, while the Mg center
has no effect at all.

To confirm this result, we place the undistorted CO$_2$ molecule at the
same position and angle of the adsorption site, with the position of the
oxygen atom near the metal center fixed, while the positions of the
carbon and the other oxygen atom shifted slightly along the line to
achieve the free CO$_2$ geometry. The frequency shift of the CO$_2$
asymmetric stretch thus results mostly from the direct effect of the
metal center.  With this geometry, our calculation shows that, while the
asymmetric stretch of CO$_2$ in Zn-MOF74 is shifted by approximately
--5.0~cm$^{-1}$, the one in Mg-MOF74 has a negligible shift of about
--0.6~cm$^{-1}$, confirming our previous results. In other words, the
fact that the oxygen atom in CO$_2$ is being close to the open metal
center has a significantly different effect depending on the metal
atoms. While the Zn atom affects the frequency strongly, the Mg atom has
almost no effect at all.  This result is quite striking, since the metal
centers Mg and Zn have a very similar valence electronic structure with
3$s$ and 4$s$ electrons as the outermost valence states. The result thus
shows that fully occupied semicore 3$d$ electrons in Zn have an important
effect in the interaction with the absorbed CO$_2$ molecules. In fact,
in Co-MOF74 and Ni-MOF74 where 3$d$ electrons are also present in the
metal center, experimentally we observed that the asymmetric stretch of
CO$_2$ has been similarly red-shifted by --10~cm$^{-1}$ as that in
Zn-MOF74, as summarized in Table~\ref{frequency} , indicating that the
3$d$ orbitals are indeed playing a similar role for the CO$_2$ frequencies
in these MOF74 systems.

The contributions of these three effects to the $\nu_3$ frequency shifts
are summarized in Table~\ref{shift}.

\begin{table}
\caption{\label{shift}Frequency shifts of $\nu_3$ (cm$^{-1}$) for
different geometries. See main text for details about these geometry.}
\begin{tabular*}{\columnwidth}{@{\extracolsep{\fill}}lcr@{}}\hline\hline
effect       & Mg-MOF74 & Zn-MOF74 \\\hline
length       & --1.6    & --3.7    \\
length+asym  & --0.5    & --3.0    \\
length+metal & --1.7    & --8.7    \\
metal        & --0.6    & --5.0    \\\hline
overall      & --0.5    & --8.1    \\
\hline\hline
\end{tabular*}
\end{table}

%%%%%%%%%%%%%%%%%%%%%%%%%%%%%%%%%%%%%%%%%%%%%%%%%%%%%%%%%%%%%%%%%%%%%%%%%
\subsection{Microscopic insight and implication of bonding}

\begin{figure}[t]
\centering
\includegraphics[width=0.8\columnwidth]{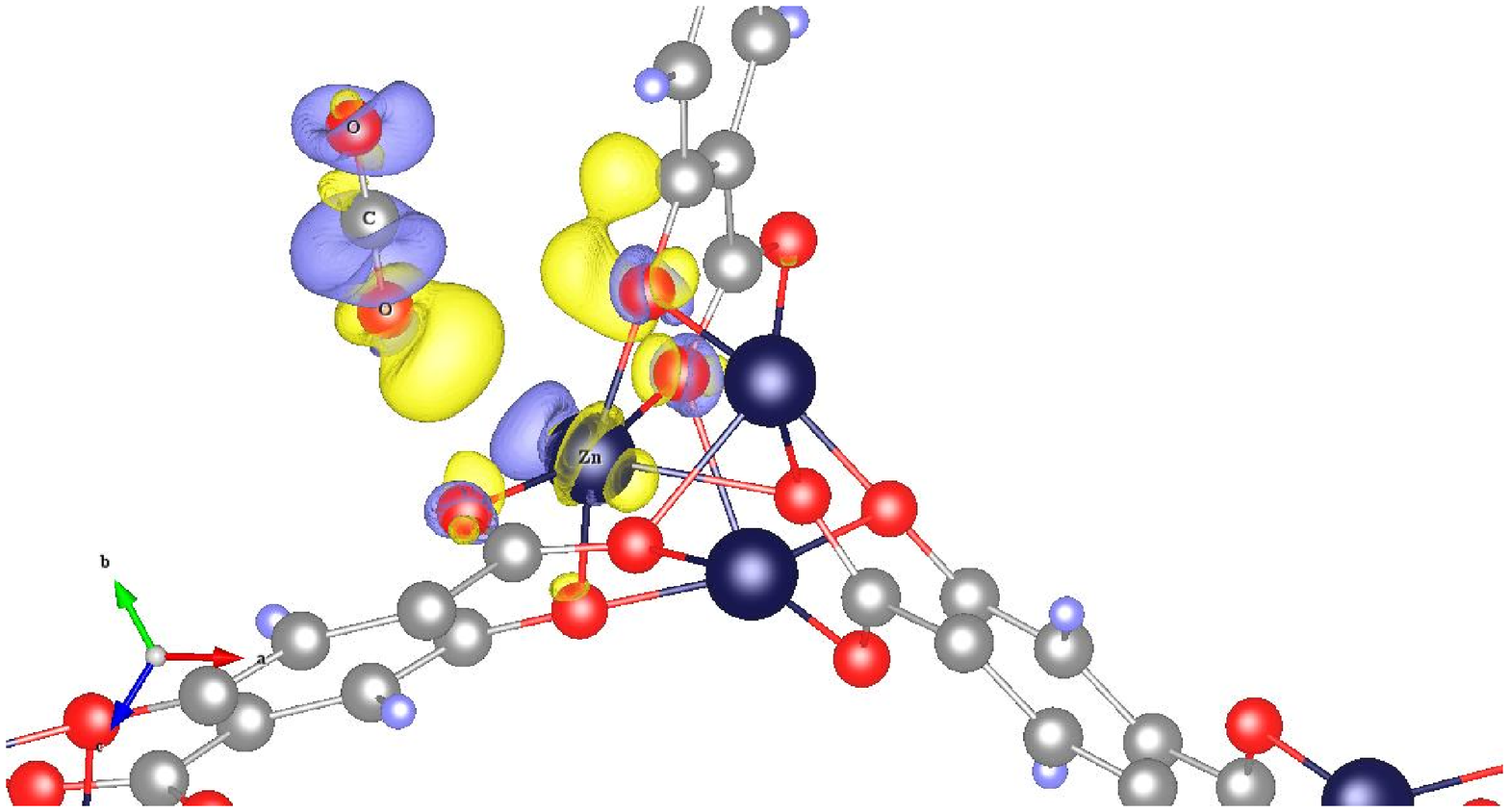}\\[4ex]
\includegraphics[width=0.8\columnwidth]{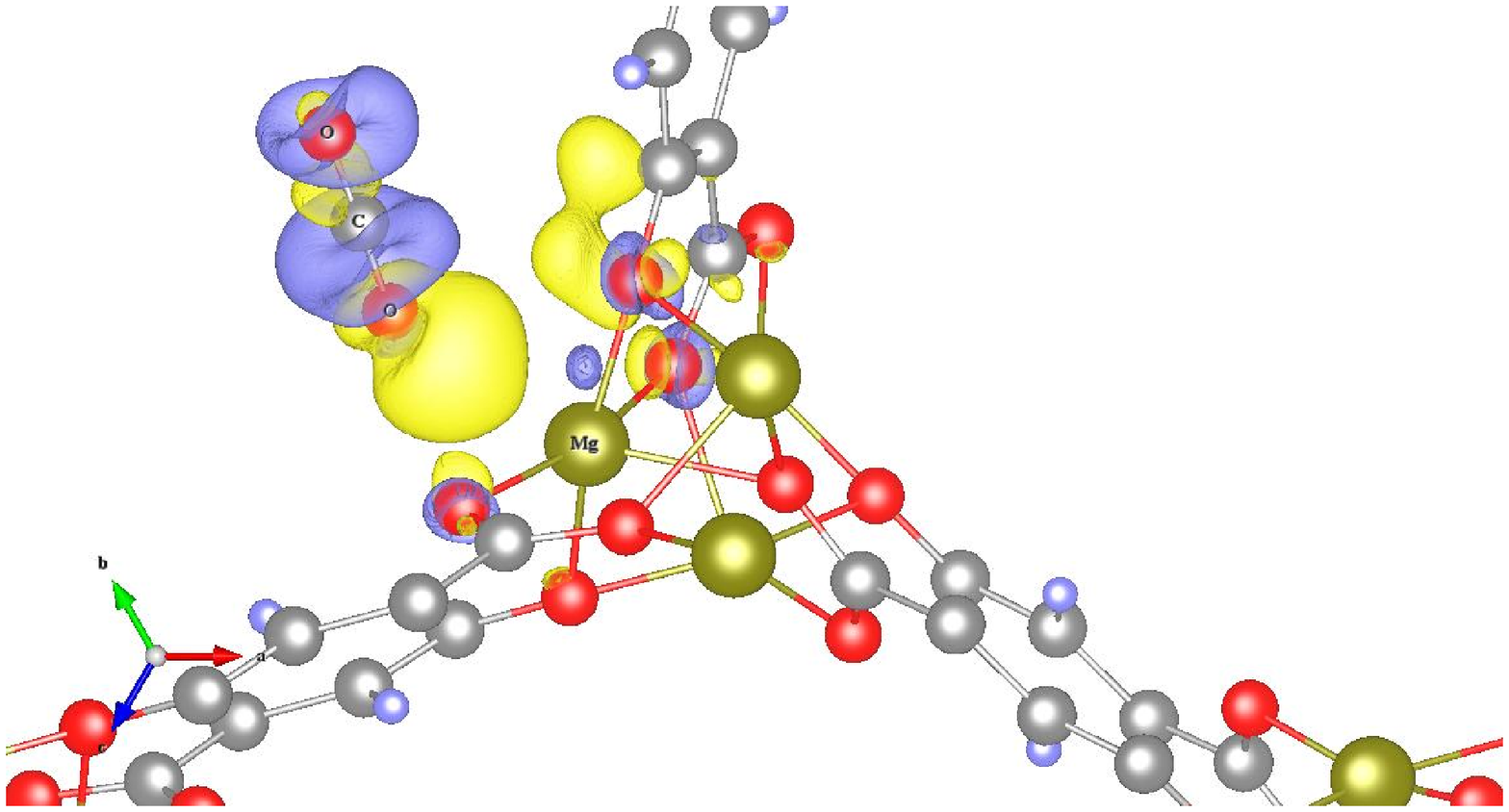}
\caption{\label{Den}(Color online)Charge density differences $\Delta\rho=\rho_{MOF74+CO_2}-\rho_{MOF74}-\rho_{CO_2}$ from before and after CO$_2$ adsorbed into Zn-MOF74 (top) and in Mg-MOF74 (bottom).}
\end{figure}

In order to gain more insight understanding for the differences in the interaction between CO$_2$ and the two different MOF74s, we investigate the charge density differences before and after CO$_2$ adsorbed inside the MOF74.  To do this, we calculate the charge densities of the CO$_2$ and MOF74 separately, by removing CO$_2$ from the MOF74, while keeping the atomic positions.  The charge density difference is defined by
\begin{equation*}
\Delta\rho=\rho_{MOF74+CO_2}-\rho_{MOF74}-\rho_{CO_2}.
\end{equation*}
The charge density difference $\Delta\rho$ thus illustrates the effect of the interaction by placing the CO$_2$ inside the MOF74.  The obtained charge density differences are plotted in Fig. \ref{Den}.

Not surprisingly, by introducing CO$_2$ inside the MOF74, the electrons in CO$_2$ are attracted toward the metal side, featuring electron deficiency in the far end (light blue color) and electron gaining near the metal site (yellow color).  This can be understood via the attraction from the positively charged metal centers, similar for both Mg and Zn centers.  However, one can also observe differences between the Zn-MOF74 and Mg-MOF74 systems, where the electron transfers more in the later than the former.  As we look at the metal center, we can observe that near Zn atom, the electrons shift slightly away from the CO$_2$ adsorption position.  This charge density change near the Zn centers are missing from those Mg center, clearly showing that it is an effect of the $d$-orbitals of the Zn atoms.  It is thus clear that the presence of the $d$-orbitals prevents the charge density transfer within the adsorbed CO$_2$ molecules, leading to smaller charge transfer as compared with those near Mg center.  As a result, the bonding energy between CO$_2$ and Zn-MOF74 is smaller than that of CO$_2$ in Mg-MOF74.  On the other hand, this clear effect of the Zn $d$-orbital effectively affect the IR frequency shift of the adsorbed CO$_2$ molecule, leading to the differences between the two systems as we discussed above.

%%%%%%%%%%%%%%%%%%%%%%%%%%%%%%%%%%%%%%%%%%%%%%%%%%%%%%%%%%%%%%%%%%%%%%%%
\subsection{Discussion}

From the previous analysis, it is now clear that the $\nu_3$ frequency
shifts of the CO$_2$ absorbed in MOF74 can be understood by the three
distinct contributions, namely, the change of the molecule length, the
off-center asymmetric distortion, and the direct effect of the open
metal site.  The interaction between CO$_2$ and the MOF causes
distortions in the molecule geometry, which affect the vibration modes.
The absorbed CO$_2$ molecules experience a stronger asymmetric
distortion in Mg-MOF74, as shown in Table~\ref{binding}, which is
consistent with the larger bonding energy calculated, as well as the
larger integrated area for the asymmetric stretch IR peak. However, such
an asymmetric distortion only has a small effect on the $\nu_3$
frequency shifts.  On the other hand, although CO$_2$ is less
asymmetrically distorted in Zn-MOF74, it experiences a larger elongation
in the overall length, which affects the $\nu_3$ frequency shift more
strongly. In addition, the nearby open metal site can play a quite
different role in affecting the CO$_2$ vibrations, where in this work,
the Zn affects strongly while Mg has barely any effect on the $\nu_3$
frequency shift.  The differences between these two MOF74 can be intuitively understood by the presence of the $d$-orbitals within the Zn atoms and the missing of those within Mg centers, as illustrated in the charge density difference plots.

The results of this work provide insight to the factors that determine
the frequency shifts of the absorbed CO$_2$ in MOF, helping to
understand the puzzling frequency shifts observed experimentally.  More
importantly, the analysis method of this work can serve as a new way to
understand the more widely examined molecule-MOF interactions and
frequency shifts. However, one must keep in mind that frequency shifts
obtained through such fixed geometries and environments reflect the
influence of different factors on the force matrix and can only give an
estimation of the influence of certain factors. In reality, originating
in the molecule-MOF interaction, all three factors are closely connected
intrinsically and it is impossible to exactly separate these different
effects.

%%%%%%%%%%%%%%%%%%%%%%%%%%%%%%%%%%%%%%%%%%%%%%%%%%%%%%%%%%%%%%%%%%%%%%%%
\section{Summary}
%%%%%%%%%%%%%%%%%%%%%%%%%%%%%%%%%%%%%%%%%%%%%%%%%%%%%%%%%%%%%%%%%%%%%%%%

In this work, we analyzed the physics determining the asymmetric
frequency shift of the CO$_2$ molecules physiadsorbed in MOFs.  Our
specific findings are summarized as follows: (i)first-principles vdW-DF2 simulations determine that the CO$_2$'s have a closer distance to the Mg center and a larger binding energy within Mg-MOF74 comparing with those in Zn-MOF74. (ii) Contrary to our intuition,  and despite the isostructure and the similarity of the open metals Mg and Zn, the
asymmetric stretch frequency of physiadsorbed CO$_2$ has been shifted stronger in Zn-MOF74 (--11~cm$^{-1}$ by IR and --8.1~cm$^{-1}$ by theory) than that in Mg-MOF74 (\mbox{$+3$~cm$^{-1}$} by IR and --0.5~cm$^{-1}$ by theory) .(iii) By comparing the response in two isostructure MOFs,
namely Zn-MOF74 and Mg-MOF74, we identified the three most important
factors contributing to the frequency shifts: the elongated CO$_2$
molecule, the off-center asymmetric distortion of the carbon atoms, and
the effect of the metal center. (iv) The asymmetric stretch frequency
is very sensitive to the overall length of the CO$_2$ molecule. Absorbed
in the MOF, the CO$_2$ molecules are elongated, which leads to a
redshift in the frequency.  This elongation effect and resulting
redshift are more significant for CO$_2$ absorbed in Zn-MOF74 compared
with those in Mg-MOF74. (v) The slight off-center asymmetric
distortion, on the other hand, favors the asymmetric stretch and causes
a slight blueshift in the frequency. (vi) Aside from changing the
geometries of the CO$_2$ molecule (i.e.\ elongating the molecule,
causing off-center asymmetric distortion of the carbon atom) and
depending on the species of the open metal site, the direct interaction
of the oxygen atom with the metal center can have very different effects
on the frequency of the asymmetric stretch, where the Zn center leads to
a redshift of about --5~cm$^{-1}$ and the Mg center has a negligible
effect on the frequency. (vii) The observed different effects of the Zn-MOF74 and Mg-MOF74 can be understood by the presence of $d$-orbital electrons in the Zn-MOF74.

%%%%%%%%%%%%%%%%%%%%%%%%%%%%%%%%%%%%%%%%%%%%%%%%%%%%%%%%%%%%%%%%%%%%%%%%
\section*{Acknowledgments}
%%%%%%%%%%%%%%%%%%%%%%%%%%%%%%%%%%%%%%%%%%%%%%%%%%%%%%%%%%%%%%%%%%%%%%%%
We would like to thank David Langreth, the father of vdW-DF, for his inspirational research. We thank
Professors \ K.\ Rabe and D.\ Vanderbilt for very helpful discussions
throughout the whole project. This work was supported in full by the
Department of Energy Grant, Office of Basic Energy Sciences, Materials
Sciences and Engineering Division, Grant No. DE-FG02-08ER46491.

%%%%%%%%%%%%%%%%%%%%%%%%%%%%%%%%%%%%%%%%%%%%%%%%%%%%%%%%%%%%%%%%%%%%%%%%


\begin{thebibliography}{99}
%%%%%%%%%%%%%%%%%%%%%%%%%%%%%%%%%%%%%%%%%%%%%%%%%%%%%%%%%%%%%%%%%%%%%%%%

\bibitem{Zeolitic}R. Banerjee, A. Phan, B. Wang, C. Knobler, H.
Furukawa, M. \'OKeeffe, and O. M. Yaghi, Science \textbf{319}, 939
(2008).

\bibitem{Morris} R. E. Morris, Nat. Chem. \textbf{3}, 347 (2011).

\bibitem{Sholl} D. S. Sholl, Nature Chemistry \textbf{3}, 429 (2011).

\bibitem{Rosi} N. L. Rosi, J. Kim, M. Eddaoudi, B. Chen, M. \'OKeeffe,
and O. M. Yaghi, J. Am. Chem. Soc. \textbf{127}, 1504 (2005).

\bibitem{Rowsell} J. L. C. Rowsell and O. M. Yaghi, J. Am. Chem. Soc.
\textbf{128}, 1304 (2006).

\bibitem{YunLiu}Y. Liu, H. Kabbour, C. M. Brown, D. A. Neumann, and C.
C. Ahn, Langmuir \textbf{24}, 4772 (2008).

\bibitem{Caskey} S. R. Caskey, A. G. Wong-Foy, and A. J. Matzger, J. Am.
Chem. Soc. \textbf{130}, 10870 (2008).

\bibitem{HWuPCL10} H. Wu, J. M. Simmons, G. Srinivas, W. Zhou, and T.
Yildirim, J. Phys. Chem. Lett. \textbf{1}, 1946 (2010).

\bibitem{Nour_H2inter} N. Nour, J. F. Veyan, L. Kong, H. Wu, Y. Zhao, J.
Li, D. C. Langreth, and Y. J. Chabal, J. Am. Chem. Soc.  \textbf{132},
14834 (2010).

\bibitem{Valenzano} L. Valenzano, B. Civalleri, S. Chavan, G. T.
Palomino, C. O. Are\'an, and S. Bordiga, J. Phys. Chem. C \textbf{114},
11185 (2010).

\bibitem{bpdc} N. Nour, P. Thissen, Y. Yao, R. C. Longo, K. Roodenko, H.
Wu, Y. Zhao, K. Cho, J. Li, D. C. Langreth, and Y. J. Chabal, J. Am.
Chem. Soc. \textbf{133}, 12849 (2011).

\bibitem{Kong_rvH} L. Kong, Y. J. Chabal, and D. C. Langreth, Phys. Rev.
B \textbf{83}, 121402(R) (2011).

\bibitem{Kong_ted} L. Kong, V. R. Cooper, N. Nijem, K. Li, J. Li, Y. J.
Chabal, and D. C. Langreth, Phys. Rev. B \textbf{79}, 081407(R) (2009).

\bibitem{HWuMetal} H. Wu, W. Zhou, and T. Yildirim, J. Am. Chem. Soc.
\textbf{131}, 4995 (2009).

\bibitem{WZhouMetal} W. Zhou and T. Yildirim, J. Phys. Chem. C
\textbf{112}, 8132 (2008).

\bibitem{XiangMe} S. Xiang, W. Zhou, Z. Zhang, M. A. Green, Y. Liu, and
B. Chen, Angew. Chem. \textbf{122}, 4719 (2010).

\bibitem{SunMetal}Y. Y. Sun, Y. Kim, and S. B. Zhang, J. Am. Chem. Soc.
\textbf{129}, 12606 (2007).

\bibitem{HoffmannBreath} H. C. Hoffmann, B. Assfour, F. Epperlein, N.
Klein, S. Paasch, I. Senkovska, S. Kaskel, G. Seifert, and E. Brunner,
J. Am. Chem. Soc. \textbf{133}, 8681 (2011).

\bibitem{SumidaMgMOF74} K. Sumida, C. M. Brown, Z. R. Herm, S. Chavan,
S. Bordiga, and J. R. Long, Chem. Commun. \textbf{47}, 1157 (2011).

\bibitem{DietzelNimof74} P. D. C. Dietzel, R. E. Johnsen, H. Fjellv\r
ag, S. Bordiga, E. Groppo, S. Chavan, and R. Blom, Chem. Commun.
\textbf{41}, 5125 (2008).

\bibitem{ChavanNiMOF74} S. Chaven, F. Bonino, J. G. Vitillo, E. Groppo,
C. Lamberti, P. D. C. Dietzel, A. Zecchina, and S. Bordiga, Phys. Chem.
Chem. Phys. \textbf{11}, 9811 (2009).

\bibitem{FitzZnmof74} S. A. FitzGerald, J. Hopkins, B. Burkholder, M.
Friedman, and J. L. C. Rowsell, Phys. Rev. B \textbf{81}, 104305 (2010).

\bibitem{synth1} N. Nijem, L. Kong, Y. Zhao, H. Wu, J. Li, D. C.
Langreth, and Y. J. Chabal, J. Am. Chem. Soc. \textbf{133} 4782 (2011).

\bibitem{FitzMOF5} S. A. FitzGerald, K. Allen, P. Landerman, J. Hopkins,
J. Matters, R. Myers, and J. L. C. Rowsell, Phys. Rev. B. \textbf{77},
224301 (2008).

\bibitem{Mgcell}H. Wu, W. Zhou and T. Yildirim, J. Am. Chem. Soc.
\textbf{131}, 4995 (2009)

\bibitem{WeiZ} W. Zhou, H. Wu, and T. Yildirim, J. Am. Chem. Soc.
\textbf{130}, 15268 (2008).

\bibitem{HughesMOF5} J. T. Hughes and A. Navrotsky, J. Am. Chem. Soc.
\textbf{133}, 9184 (2011).

%===================================================synthes
\bibitem{synth3} P. D. C. Dietzel and R. Blom, H. Fjellvag, Eur. J.
Inorg.  Chem \textbf{2008}, 3624 (2008).

\bibitem{synth4} P. D. C. Dietzel, R. E. Johnsen, R. Blom, and H.
Fjellvag, Chem. Eur. J. \textbf{14}, 2389 (2008).

\bibitem{synth5} P. D. C. Dietzel, Y. Morita, R. Blom, and H. Fjellvag,
Angew. Chem. Int. Ed. \textbf{44}, 6354 (2005).
%===================================================================

%##########################################method
\bibitem{vdW-DF2} K. Lee, \'E. D. Murray, L. Kong, B. I. Lundqvist, and
D. C. Langreth, Phys. Rev. B \textbf{82}, 081101(R) (2010).

\bibitem{vdW-DF} M. Dion, H. Rydberg, E. Schr\"oder, D. C. Langreth, and
B. I. Lundqvist, Phys. Rev. Lett. {\bf 92}, 246401 (2004); {\bf 95},
109902 (2005).

\bibitem{potential_PRB} T. Thonhauser, V. R. Cooper, Li Shen, A. Puzder,
P. Hyldgaard, and D. C. Langreth,  Phys. Rev. B {\bf 76}, 125112 (2007).

%======================================vdwDFapplication
\bibitem{Langreth_rev} D. C. Langreth et al., J. Phys.: Condens. Mattter
{\bf 21}, 084203 (2009).

\bibitem{dimers} T. Thonhauser, A. Puzder, and D.C. Langreth, J. Chem.
Phys. {\bf 124}, 164106 (2006); S. Li, V.R. Cooper, T. Thonhauser, A.
Puzder, and D. C. Langreth, J. Phys. Chem. A {\bf 112}, 9031 (2008); J.
Hooper, V. R. Cooper, T. Thonhauser, N. A. Romero, F. Zerilli, and D. C.
Langreth, ChemPhysChem {\bf 9}, 891 (2008).

\bibitem{physisorption} M. Mura, A. Gulans, T. Thonhauser, and L.
Kantorovich, Phys. Chem. Chem. Phys. {\bf 12}, 4759 (2010).

\bibitem{drug} V. R. Cooper, T. Thonhauser, A. Puzder, E. Schr\"{o}der,
B. I. Lundqvist, and D. C. Langreth, J. Am. Chem. Soc. {\bf 130}, 1304
(2008); V. R. Cooper, T. Thonhauser, and D. C. Langreth, J. Chem. Phys.
{\bf 128}, 204102 (2008); S. Li, V. R. Cooper, T. Thonhauser, B. I.
Lundqvist, and D. C. Langreth, J. Phys. Chem. B {\bf 113}, 11166 (2009).

%#######################################################################ABINIT
\bibitem{ABINIT1} X. Gonze, B. Amadon, P.-M. Anglade, J.-M. Beuken, F.
Bottin, P. Boulanger, F. Bruneval,' D. Caliste, R. Caracas, M. Cote, T.
Deutsch, L. Genovese, Ph. Ghosez, M. Giantomassi, S. Goedecker, D.R.
Hamann, P. Hermet, F. Jollet, G. Jomard, S. Leroux, M. Mancini, S.
Mazevet, M.J.T. Oliveira, G. Onida, Y. Pouillon, T. Rangel, G.-M.
Rignanese, D. Sangalli, R. Shaltaf, M. Torrent, M.J. Verstraete, G.
Zerah, J.W. Zwanziger, Computer Phys. Commun. \textbf{180}, 2582 (2009).

\bibitem{ABINIT2}X. Gonze, G.-M. Rignanese, M. Verstraete, J.-M. Beuken,
Y. Pouillon, R. Caracas, F. Jollet, M. Torrent, G. Zerah, M. Mikami, Ph.
Ghosez, M. Veithen, J.-Y. Raty, V. Olevano, F. Bruneval, L. Reining, R.
Godby, G. Onida, D.R. Hamann, and D. C. Allan, Zeit. Kristallogr.
\textbf{220}, 558 (2005).
%======================================================================

\bibitem{Troulier} N. Troullier and J. L. Martins, Phys. Rev. B
\textbf{43}, 1993 (1991).
%===============================================

\bibitem{NIST} NIST Standard Reference Database 69: Nist Chemistry
WebBook at  http://webbook.nist.gov/chemistry.


\end{thebibliography}
\end{document}